\documentstyle[12pt,epsfig]{article}
\def\requals{\mathrel{\raise2pt\hbox to 8pt{\raise -7pt\hbox{$=$}\hss{$>$}}}}

\begin{document}

\begin{center}

{\Large {\bf Dimensional Regularization and Nuclear Potentials}}\\

\vspace*{0.20in}

J.L.\ Friar \\
Theoretical Division \\
Los Alamos National Laboratory \\
Los Alamos, NM  87545 \\
\end{center}

\begin{abstract}

It is shown how nucleon-nucleon potentials can be defined in N dimensions, using
dimensional regularization to continue amplitudes.  This provides an easy way to
separate out contact ($\delta$-function) terms arising from renormalization.  An
example is worked out several ways for the case of two scalar particles
exchanged between nucleons, which  involves a very simple loop calculation. 
This leads to a Feynman-parameterized representation for the nucleon-nucleon
potential.  Alternately, a dispersion representation can be developed leading to
a different, though equivalent, form. 

\end{abstract}

It might be surmised that there exist no {\bf new} ways to develop nuclear
potentials after decades of experimentation with techniques.  We present below a
variation on several such techniques, which, to the best of our knowledge, has
not been used before.  As an illustration, we perform in several different ways
a simple calculation involving a closed meson loop, which leads to a
two-boson-exchange potential (TBEP). 

\begin{figure}[hb]
  \epsfig{file=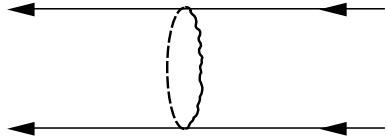,height=4.5in,angle=90}
  \caption{The nucleon-nucleon potential resulting from the
     simultaneous exchange of two scalar particles, ``a'' and ``b''.}
\end{figure}

Figure 1 shows the interaction of two nucleons ``1'' and ``2'' via the exchange
of two different scalar mesons, ``a'' and ``b''.  The corresponding Lagrangian
is 
$$
L_{INT} = \lambda \overline{N} N ab \, . \eqno (1)
$$
The nonrelativistic potential corresponding to Figure (1) can be developed in
old-fashioned second-order perturbation theory in the usual way\cite{fc} for a
nucleon-nucleon separation, r: 
$$
V_{12} (r) = -2 \lambda^2 \int \frac{d^3 q_a} {(2 \pi)^3}  
\frac{e^{i\mbox{\boldmath $q$}_a \cdot \mbox{\boldmath $r$}}}{(2E_a)} \int  
\frac{d^3 q_b}{(2 \pi)^3} \frac{e^{i \mbox{\boldmath $q$}_b \cdot  
\mbox{\boldmath $r$}}}{(2E_b) (E_a + E_b)} \, . \eqno (2)
$$ 
This awkward and inelegant expression is nevertheless very easy to interpret.
The initial factor of two accounts for two ways the mesons can be exchanged:
``1'' to ``2'' or from ``2'' to ``1''.  The factors $(2E_a)$ and $(2E_b)$ are
the wave function normalization factors for ``a'' and ``b'' $(E_x = (q^2_x +
m^2_x)^{\frac{1}{2}})$, while $-(E_a + E_b)$ is the energy denominator
(neglecting nucleon recoil) and the integration is over the mesons' phase
spaces.  This can be cast into a more conventional form by using\cite{rij}
$$
\frac{1}{E_a E_b (E_a + E_b)} = \frac{2}{\pi} \int^{\infty}_{0}  
\frac{d \beta}{(E^2_a + \beta^2)(E^2_b + \beta^2)} \, , \eqno(3)
$$
which allows a convolution representation to be written in terms of separate
Yukawa functions for a and b exchange: 
$$
V_{12} (r) = \frac{-4 \lambda^2}{(4 \pi)^3 r^2} \int^{\infty}_{0} d{\beta} \,
e^{-\left( \sqrt{\beta^2 + m^2_a} + \sqrt{\beta^2 + m^2_b}\right) r} \eqno (4a)
$$
$$
= \frac{-2 \lambda^2}{(4 \pi)^3 r^2} \int^{\infty}_{m_{a} + m_{b}}  
\frac{dy \, e^{-yr}}{y^2} 
\frac{(y^2 + m^2_a - m^2_b)(y^2 + m^2_b - m^2_a)}
{[y^4 - 2 y^2 (m^2_a + m^2_b) + (m^2_a - m^2_b)^2 ]^{\frac{1}{2}}} \eqno (4b)
$$
$$
= \frac{-2 \lambda^2}{(4 \pi)^3 r} \int^{\infty}_{m_{a} + m_{b}} \frac{dy}{y}
e^{-yr} [y^4 - 2y^2 (m^2_a + m^2_b) + (m^2_a - m^2_b)^2]^{\frac{1}{2}} \, .
\eqno (4c)
$$
We have made a change of variables to obtain eq.\ (4b) and an integration by
parts to realize eq.\ (4c).  Our points can be made if we further assume that
one particle mass vanishes or that $m_a = m_b = m$; the latter leads to an
elementary integral which we evaluate: 
$$
V_{12}(r) = \frac{-2 \lambda^2}{(4 \pi)^3 r^2} \int^{\infty}_{2m} dy\, e^{-yr}  
\frac{y}{(y^2 - 4m^2)^{\frac{1}{2}}} \eqno (5a)
$$
$$
= \frac{-4m \lambda^2}{(4 \pi)^3 \, r^2} K_1 (2mr) \, . \eqno (5b)
$$
Note the factor of $(4 \pi)^3$, one power from the configuration space 
propagator and two powers from the loop integral. Both are necessary for
dimensional power counting to work\cite{geo}.

On the other hand, we can calculate the Feynman diagram associated with the loop
in Figure (1).  It is easy to show\cite{bd} that the potential equivalent to
that amplitude, M, should have an additional factor of $i$:  $V = iM$. 
Evaluating the diagram we find 
$$
M(q) = (\overline{N}_1 N_1)(\overline{N}_2 N_2) \lambda^2 I(q^2) \, , \eqno (6a)
$$
where $q$ is the transferred momentum. The divergent loop integral is given by
$$
I(q^2)=\int \frac{d^4 k}{(2 \pi)^4} \frac{1}{(k^2 - m^2_a)((q - k)^2 - m^2_b)}\, ,
\eqno (6b)
$$
and dropping the unimportant nucleon spinor factors we find
$$
V(q^2) = i \lambda^2 I(q^2) \, . \eqno (6c)
$$
Evaluating divergent loop integrals requires regularization, and for a variety
of reasons, dimensional regularization is the method of choice\cite{th} today. 
This yields 
$$
I_N (q^2) \rightarrow \mu^{4 - N} \int \frac{d^N k}{(2 \pi)^N} 
\frac{1}{(k^2 - m^2_a)} \frac{\;\;}{((q - k)^2 - m^2_b)} \, , \eqno (6d)
$$
where the number of dimensions has been extended from 4 to N, and for N $<$ 4
the integral is convergent.  The renormalization scale $\mu$ keeps the overall
dimensionality of the integral the same. 

Using the Feynman parameterization\cite{bd}
$$
\frac{1}{ab} = \int^1_0 \frac{dz}{[a + (b - a)z]^2} \, , \eqno(7)
$$
and shifting the variable to $k^{\prime} = k - q z$ leads to
$$
I_N (q^2) = \int^1_0 dz \int \frac{d^N k^{\prime}}{(2 \pi)^N} 
\frac{1}{[k^{\prime 2} - (m^2_a  - (m^2_a - m^2_b)z - q^2 z (1 - z))]^2} 
\eqno (8a)
$$
$$
= \frac{i}{(4 \pi)^2} (4 \pi \mu^2)^{2 - \frac{N}{2}} \, \Gamma (2 - \frac{N}{2})  
\int^1_0 d z \, [m^2_a - (m^2_a - m^2_b) z - q^2 z (1 - z)]^{\frac{N}{2} - 2} 
\, . \eqno (8b)
$$
The latter result is a standard form for one-loop amplitudes\cite{dgh}. Usually
at this point one writes $4 - N = \epsilon$ and performs an
$\epsilon$-expansion, leading to a divergent constant (which generates a contact
term, $(\overline{N} N)^2$, in the nucleon-nucleon force) and a finite
logarithm.  We eschew this approach and define a potential in $N$ space-time or
n space dimensions $(n = N-1)$: 
$$
V_N (r) \equiv  \int \frac{d^n q}{(2 \pi)^n} e^{i \mbox{\boldmath  
$q$} \cdot \mbox{\boldmath $r$}} [i \lambda^2 I_N (q^2)] \, , \eqno (9)
$$
where we choose to work in the frame where $q_0 = 0$ ($I_N$ is a function of
$q^2 = q^2_0 - \mbox{\boldmath $q$}^2)$.  Alternatively, the energy transfer,
$q_0$, is small and of order $(\frac{v}{c})^2$ (i.e., a relativistic correction)
and can be dropped.  Both $\mbox{\boldmath $q$}$ and $\mbox{\boldmath $r$}$ are
vectors in $n$ space dimensions.  The angular integrals can be
evaluated\cite{th} which leads to 
$$
V_N (r) = \frac{r^{1 - \frac{n}{2}}}{(2 \pi)^{\frac{n}{2}}} \int^{\infty}_{0} d
q\, q^{\frac{n}{2}} J_{\frac{n}{2} - 1} (qr) \left[i \lambda^2 I_N (-q^2) 
\right] \, , \eqno (10) 
$$
which can be verified easily for $n = 3$. Note that we are using $q^2 \equiv
\mbox{\boldmath $q$}^2$ as an integration variable, which accounts for the sign
change in the argument of $I_N$. Inserting expression (8b) for $I_N$ 
and performing the $q$-integral leads to: 
$$
V_N (r) = \frac{-(4 \pi \mu^2)^{2 - \frac{N}{2}} \, \lambda^2}{\sqrt{2} 
(4 \pi)^2 \pi^{\frac{n}{2}}\, r^{n - \frac{3}{2}}} \int^1_0 \frac{d z \, 
\beta^{n - \frac{3}{2}} \, K_{n - \frac{3}{2}} (\beta r)}{[z (1 - z)]^{2 - 
\frac{N}{2}}} \, , \eqno (11a)
$$
where
$$
\beta^2 (z) = \frac{m^2_a - (m^2_a - m^2_b)z}{z(1 - z)} > 0\,  . \eqno (11b)
$$
The factors of 2 and $\pi$ clearly depend on $n$ (or $N$). Note that the
divergent (for $N = 4$) $\Gamma$-function has disappeared and the result is
finite for $r \neq 0$.  Standard Bessel function identities\cite{as} allow us to
write for $N = 4$: 
$$
V_4 = \frac{-2 \lambda^2}{(4 \pi)^3 r^3} \int^1_0 d z \, e^{-\beta r} 
(1 + \beta r) \, .\eqno (12)
$$
For $m_a = m_b$ this can be shown\cite{gr} to equal eq.\ (5b).  Note that this
representation for the potential involves the Feynman parameterization variable,
$z$.  Our derivation (12) is both elegant and more directly related to the
Feynman loop diagrams than is the conventional derivation. Indeed, dimensional
regularization was developed for these diagrams\cite{th} 

The reason why the divergent factor $\Gamma (2 - \frac{N}{2})$ disappears is
that we implicitly renormalized the loop graph when we developed the potential.
Performing an $\epsilon$-expansion $(\Gamma (2 - \frac{N}{2}) \sim
\frac{2}{\epsilon} +$ finite) in eq.\ (8b) leads to a contact term $( \sim
\delta^3 (\mbox{\boldmath $r$}))$.  By keeping the internucleon separation, 
$r$, finite, these terms don't arise.  They have been ``regularized'' away.  
Note that $V_N (r) \sim \frac{1}{r^{2N - 5}}$ for small $r$ and is 
progressively more singular for increasing $N$. 

\begin{figure}[htb]
\epsfig{file=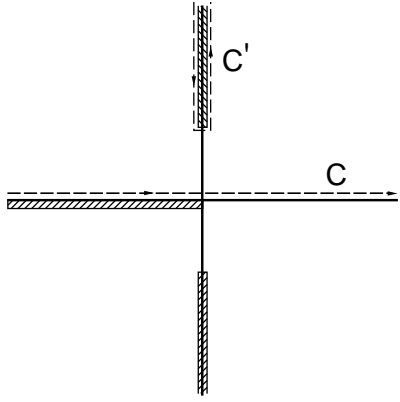,height=2.5in,bbllx=-35pt,bblly=339pt,bburx=438pt,bbury=630pt}
  \caption{The analytic q-plane illustrating branch cuts discussed
        in the text.  The integration contour-C, which defines the 
        Fourier transform, can be deformed to C$^{\prime}$, which 
        gives the dispersion representation for the potential.}
\end{figure}

We can derive this result and an alternative representation (equivalent to eq.\
(4c), for $N = 4$) by using the analytic properties of the field-theoretic
amplitude.  In the t-channel of Fig.\ (1) the reaction $\overline{N} + N
\rightarrow a + b$ is open, and this information allows construction of a
potential, an old but very useful technique\cite{bj}. The amplitude $I_N (q^2)$
in eq.\ (9) is real for $q^2 < 0$, but for $q^2 >0$ it develops an imaginary
part above the threshold for the reaction, which requires $q^2 \geq (m_a +
m_b)$. Switching to the integration variable $q^2 \equiv \mbox{\boldmath $q$}^2$
used in eq. (10), this implies branch cuts for $q^2 < 0$ or along the imaginary
axis in the complex $q$-plane indicated in Fig.\ (2).  The left-hand cut arises
from $q^{\frac{n}{2}} J_{\frac{n}{2} -1} (qx)$.  Writing $J_{\nu}(z) =
\frac{1}{2} \left[H^{(1)}_{\nu} (z) + H^{(2)}_{\nu} (z) \right]$ and noting that
$H^{(1)}_{\nu}$ behaves asymptotically as $e^{iz}$ while $H^{(2)}_{\nu}$ behaves
as $e^{-iz}$, we can write 
$$
J_N = \int^{\infty}_{0} dq \, q^{\frac{n}{2}} J_{\frac{n}{2} - 1} (qr) 
I_N (-q^2) = \frac{1}{2} \int_C dq \, q^{\frac{n}{2}} H^{(1)}_{\frac{n}{2} -1} 
(qr) I_N (-q^2) \, , \eqno (13) 
$$
using $H^{(1)}_{\nu} (e^{i \pi}r) = H^{(2)}_{\nu} (r) e^{- \pi i (\nu + 1)}$.
The integral along $C$ can be continuously deformed into $C^{\prime}$ because
the integral vanishes exponentially in the upper half of the $q$-plane.  Using
the properties of the $H^{(1)}_{\nu}$ function for imaginary argument and 
$q = i x \pm \epsilon$, we then find the simple and elegant result: 
$$
J_N = \frac{2}{\pi} \int^{\infty}_{m_{a} + m_{b}} dx \, x^{\frac{n}{2}} \,
K_{\frac{n}{2} - 1} (xr) \, \Im \left[I_N (x^2 + i \epsilon) \right] \; , 
\eqno (14)
$$
which provides an excellent representation for the potential.  Note that the
range of the force varies from $(m_a + m_b)$ to $\infty$.  Putting everything
together from eqns.\ (8b) and (10), we obtain 
$$
V_N = \frac{-2 \lambda^2 (4 \pi \mu^2)^{2 - \frac{N}{2}} \, \Gamma (2 -  
\frac{N}{2})}{\pi (2 \pi)^{\frac{n}{2}} r^{\frac{n}{2} - 1} (4 \pi)^2}  
\int^{\infty}_{m_{a} + m_{b}} dx \, x^{\frac{n}{2}} \, K_{\frac{n}{2} - 1} (xr)
$$
$$
\Im \left[ \int^1_0 \frac{dz \hspace*{0.50in}}{[z(1 - z)]^{2 -
\frac{N}{2}} [\beta^2 - x^2 - i \epsilon]^{2 - \frac{N}{2}}} \right] \; , \eqno
(15a) 
$$
where $\beta^2 = [m^2_a - (m^2_a - m^2_b) z]/z(1 - z)$.  Evaluating the
imaginary part leads to 
$$
V_N = \frac{-2 \lambda^2 (4 \pi \mu^2)^{2 - \frac{N}{2}}}{(2 \pi)^{\frac{n}{2}}
(4 \pi)^2 \, \Gamma (\frac{N}{2} - 1) \, r^{\frac{n}{2} - 1}} 
\int^{\infty}_{m_{a} + m_{b}} dx \, x^{\frac{n}{2}} \, K_{\frac{n}{2} - 1} (xr)
$$
$$
\int^1_0 \frac{dz \, \theta (x^2 - \beta^2)}{[z(1 - z) ( x^2 - \beta^2 ) 
]^{2 -  \frac{N}{2}}} \; , \eqno (15b)
$$
where the singularity at $N = 4$ has disappeared.

Two options are available for further simplification.  One can first evaluate
the $x$ integral, which leads immediately to eq.\ (11a).  Alternatively, one can
perform the $z$ integral first.  The function $\beta (z)$ is larger than $x^2$
for values of $z$ near $0$ and $1$, making the $\theta$-function vanish.  The
argument of that function is positive between two values of $z \left( z_{\pm} =
\frac{x^2 + \Delta m^2}{2x^2} \pm \frac{1}{2} \left[ (1 + \right. \right.$ 
$\left. \left. \frac{\Delta m^2}{x^2})^2 - \frac{4 m^2_a}{x^2}
\right]^{\frac{1}{2}} \right)$, where $\Delta m^2 = m^2_a - m^2_b$, which resets
the limits on the integral.  The resulting integral is a beta-function [7], and
one obtains 
$$
V_N (r) = \frac{-2 \lambda^2 (4 \pi \mu^2)^{2 - N/2} \, 2^{3/2-n}}{(2
\pi)^{\frac{n - 1}{2}} \, r^{\frac{n}{2} -1} \, (4 \pi)^2 \, \Gamma
(\frac{n}{2})} \int^{\infty}_{m_{a} + m_{b}} \frac{dx}{\sqrt{x}} \,
K_{\frac{n}{2} - 1}(xr) \left[ (x^2 + \Delta m^2)^2 - 4 m^2_a x^2
\right]^{\frac{N-2}{4}} \, . \eqno (16) 
$$
This reduces to eq.\ (4c) for $N = 4$.

Equations (11) and (16) are our principal results and are equivalent.  The
former is a potential defined in $N$ dimensions which exploits the Feynman
parameterization of the loop integral.  The latter is a dispersion
representation which exploits the analytic properties of the amplitude, and was
also derived using a trick and simple, old-fashioned perturbation theory (in 4
dimensions).  Infinities which naturally arise are regularized and disappear
from the final form.  Dimensional regularization was used to define the
continuation to $N$ dimensions, and both our approach and our techniques
maintain a strong connection between the potential and the underlying field
theory. 

\begin{center}
{\bf Acknowledgements}
\end{center}

This work was performed under the auspices of the U.\ S.\ Department of Energy.

\end{document}